\documentclass[aps,pra,twocolumn,floatfix,superscriptaddress,longbibliography]{revtex4-2}

\usepackage{graphicx}
\usepackage{amssymb}
\usepackage{amsmath}
\usepackage{color}
\usepackage{psfrag}
\usepackage{epsfig}
\usepackage{bbm}
\usepackage{bm}
\usepackage{hyperref}
\usepackage[normalem]{ulem}
\usepackage{amsthm}
\usepackage{epstopdf}
\usepackage{verbatim}
\usepackage{subfigure}
\usepackage[export]{adjustbox}
\usepackage{natbib}
\usepackage{color}

\DeclareMathOperator{\tr}{tr}

\begin{document}
\title{Quantum direct cause across the Cherenkov threshold in circuit QED}

\author{Jhen-Dong Lin}
\affiliation{Department of Physics, National Cheng Kung University, 701 Tainan, Taiwan}
\affiliation{Center for Quantum Frontiers of Research \& Technology, NCKU, 70101 Tainan, Taiwan}

\author{Yueh-Nan Chen}
\email{yuehnan@mail.ncku.edu.tw}
\affiliation{Department of Physics, National Cheng Kung University, 701 Tainan, Taiwan}
\affiliation{Center for Quantum Frontiers of Research \& Technology, NCKU, 70101 Tainan, Taiwan}

\date{\today}

\begin{abstract}
We investigate Cherenkov radiation triggered by qubit acceleration, which can be simulated using superconducting circuits. By analyzing qubit dynamics, we confirm existence of the Cherenkov speed threshold. A question immediately arises: What is the role of the Cherenkov speed threshold from the aspect of causation? More specifically, what is the effect of the threshold on the ability of the qubit to transmit quantum information? To address this question, we consider measurements of the quantum direct cause, which can be used to estimate channel capacity, based on a recently developed notion on temporal quantum correlations. When choosing proper values for qubit acceleration and qubit-field coupling in a single-mode model, we surprisingly discover that the Cherenkov threshold serves as the speed limit for quantum information propagation in the single-mode model. We further extend use of these measurements to a multi-mode model. The results indicate that introducing extra modes can lead to further suppression of the quantum direct cause. The suppression is further enhanced when the number of field modes involved in the system is increased.
\end{abstract}

\maketitle

\section{Introduction}
When a charged particle is moving at a speed faster than the speed of light in a medium (Cherenkov speed threshold or, more concisely, Cherenkov threshold), the particle will start to emanate radiation. This radiation is known as the Cherenkov radiation~\cite{cherenkov1934,vavilov1934,frank1937coherent,PhysRevX.8.041013,bogdanov2012angular,koch2006cherenkov,volotka2013progress,shi2018superlight,macleod2019cherenkov,carusotto2013cerenkov,luo2003cerenkov,liu2012surface,kaminer2016efficient,kremers2009theory,ginzburg1940quantum,kaminer2016quantum,marino2017,belgiorno2010,calajo2017strong,kaminer2016quantum,marino2017,belgiorno2010,calajo2017strong}, which was first analyzed in the field of classical electromagnetism~\cite{frank1937coherent}. Quantum mechanical treatment of the Cherenkov effect, which was first proposed by Ginzburg~\cite{ginzburg1940quantum}, demonstrates that the moving source can be a neutral body or any sort of perturbation~\cite{marino2017,belgiorno2010,kaminer2016quantum}. Recent studies~\cite{sabin2017,felicetti2015,garcia2017,PhysRevA.99.052328} have shown that Cherenkov radiation can be observed in superconducting circuits via simulations of a qubit moving in constant velocity and with tunable coupling strength. Along this line of thinking, we further consider the Cherenkov effect that is triggered by uniformly accelerating motion. In this study, our goal is to reveal the role of the Cherenkov threshold from the aspect of quantum causation. To accomplish this goal, a causality test, for estimating the underlying speed of quantum information carried by the accelerating qubit, is required. 

A notable causality test usually performed on cavity and circuit QED systems~\cite{munoz2018,jonsson2014,sabin2011,zohar2011} is known to be a Fermi problem~\cite{fermi1932} and is used to estimate the propagation speed of photons in a cavity. The basic principle of the test is to consider two distantly separated atoms in the cavity. Initially, one of the atoms is prepared in the excited state and the other one in the ground state. The excited atom will decay and emit a photon, which can be received by the other atom. The speed of the photon can then be intuitively estimated based on the flying time to excite the other atom. In our study, instead of performing Fermi's test, we aim to uncover the speed of quantum information transmitted/carried by the accelerating qubit. Therefore, we employ recently developed methods for inferring what is referred to as the quantum direct cause based on pseudo-density matrix (PDM) formalism and its related idea known as temporal quantum steering.

PDM~\cite{fitzsimons2015} is established through a temporal analogue of the quantum state tomography procedure, which signifies that a PDM can be constructed via time-like separated measurements, i.e., measurements performed on the same system at different times. A PDM can be a negative matrix, which fails to be reinterpreted as a valid quantum state. The existence of negative eigenvalues naturally eliminates common-causal explanations~\cite{allen2017quantum,ried2015quantum,costa2016quantum}, indicating the presence of a quantum direct cause (or direct-causal influence) encoded in the PDM. Furthermore, recent research~\cite{PhysRevLett.123.150502} has demonstrated that the degrees of direct causal influence between two ends of a quantum channel also implies the upper bound of its channel capacity. Therefore, in this study, measuring the quantum direct cause is equivalent to determining the capability of the accelerating qubit to propagate quantum information. 

On the other hand, the notion of temporal steering (TS)~\cite{chen2014,chen2016,ku2016,bartkiewicz2016,chen2017,ku2018} was introduced in analogy with the steering concept~\cite{schrodinger1935discussion,wiseman2007steering,jones2007entanglement,cavalcanti2009experimental,piani2015necessary,skrzypczyk2014quantifying,costa2016quantification,wollmann2016observation,cavalcanti2016quantum,uola2019quantum} proposed by Schr\"{o}dinger~\cite{schrodinger1935discussion}, where remote state preparation is made possible through the use of entangled pairs. Instead of involving bipartite systems, TS explores the possibility of reformulating the steering task through a single system at different moments. A recent paper by Ku \textit{et al.}~\cite{ku2018} has further highlighted that there exists a hierarchical relation between PDM and TS, suggesting that TS is a weaker measure of the quantum direct cause.

The rest of this paper is structured as follows. In section \ref{the_model}, we begin with the single-mode model, where the Cherenkov effect can be triggered by an accelerating qubit. We then identify the Cherenkov threshold via the perturbation theory. By simulating the qubit dynamics with different initial states, we can determine that either a decrease in the qubit acceleration or an increase in the qubit-field coupling strength can cause the qubit states to approach an excited state after the threshold is crossed. It then becomes more difficult to distinguish the qubit initial condition through the qubit states, implying that modulation of the acceleration and coupling strength enhances the tendency of the qubit to lose quantum information. In section \ref{causal_measure}, we then review measurements of the quantum direct cause, which can be used to quantify the ability of the accelerating qubit to propagate quantum information. In section \ref{numerical_res}, we present numerical results regarding the quantum direct cause for the single-mode model. We observe that, if we choose proper values for the acceleration and coupling strength, the quantum direct cause can drastically drop to zero when the Cherenkov threshold is crossed. Under these circumstances, the threshold serves as a speed limit for the transmission of quantum information via the qubit. Afterward, we consider a multi-mode model for simulating a genuine cavity. The results suggest that introducing extra modes can lead to further suppression of the direct-causal influence. In addition, the suppression is further enhanced when the number of modes is increased. From the perspective of an open quantum system, the enhancement of the suppression is an effect of the increase in system--reservoir interactions, i.e., the enlarged size of the environment.

\section{The model \label{the_model}}
\begin{figure}
\includegraphics[width=1\linewidth]{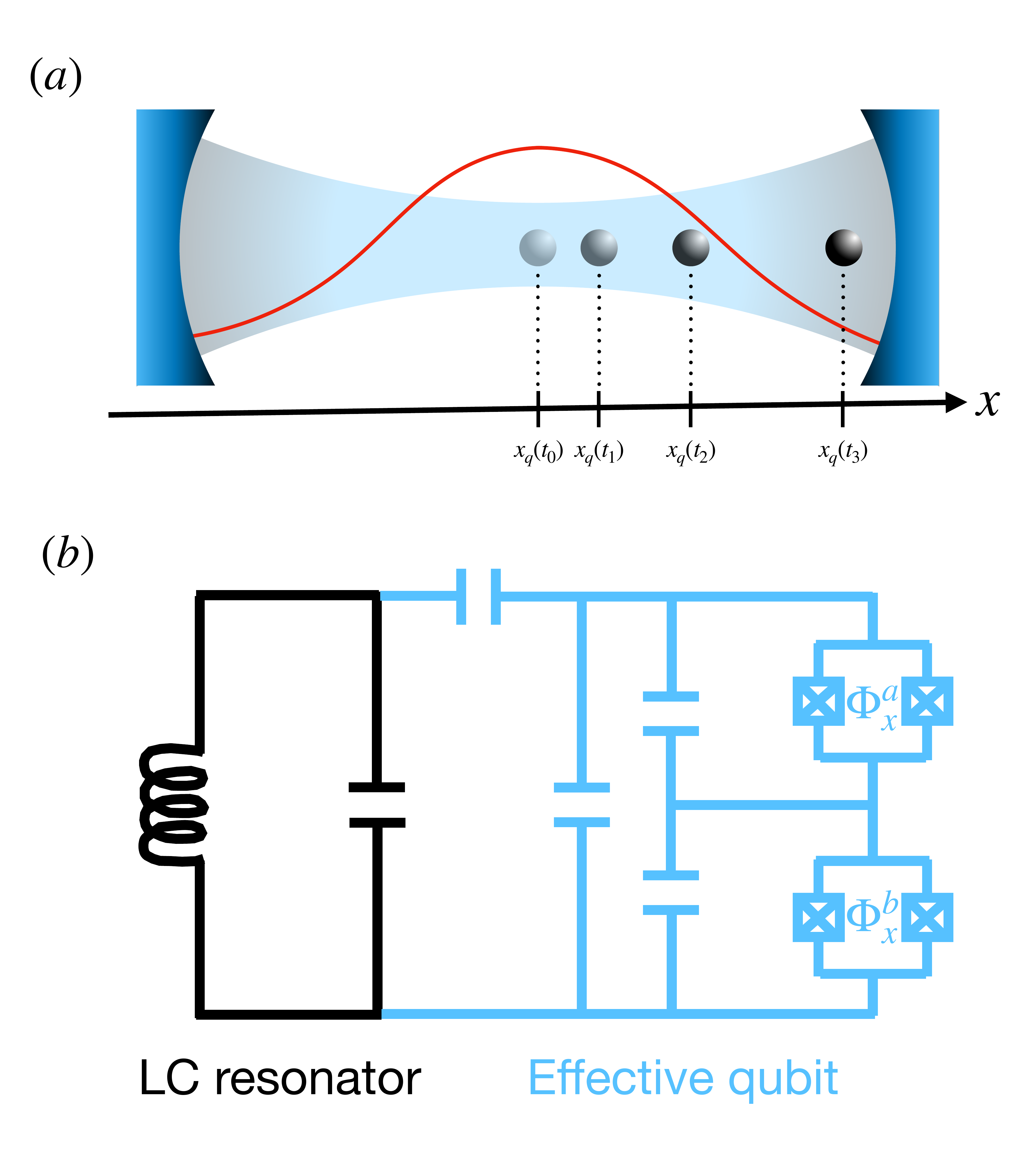}
\caption{\label{model}(a) Uniformly accelerating qubit coupled to cavity, where red curve represents spatial profile of coupling strength. (b) System can be simulated using superconducting circuit with tunable coupling strength, which depends on external fluxes $\Phi_x^a$ and $\Phi_x^b$.}
\end{figure}

\begin{figure}
\includegraphics[width=1.05\linewidth]{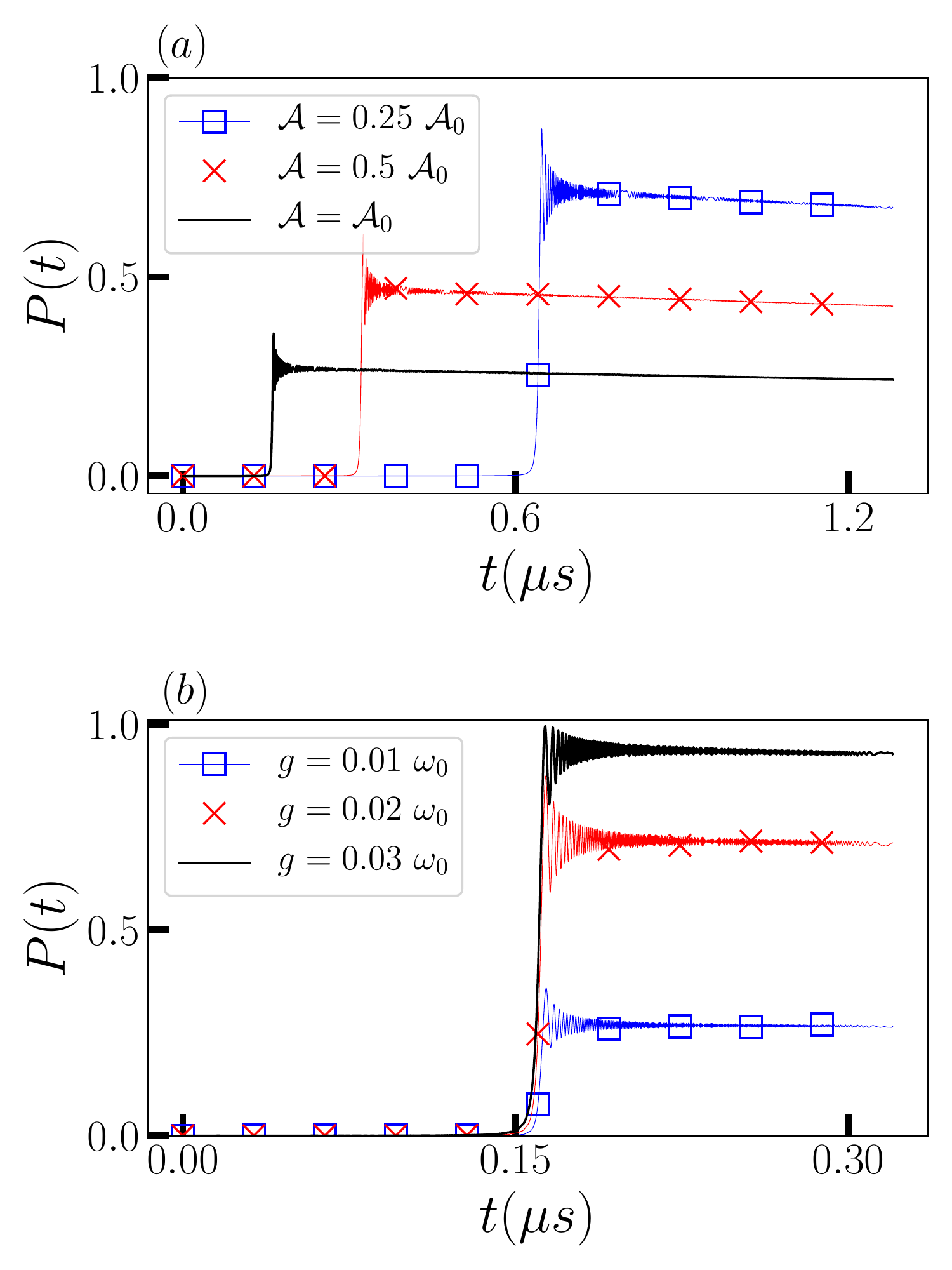}
\caption{\label{radiation_p} Numerical simulations of qubit excitation probability with initial qubit-field state $|G,0\rangle$. (a) Acceleration is tuned from $0.25\mathcal{A}_0$ to $\mathcal{A}_0$ with $\mathcal{A}_0 = 10^{15}m/s^2$, and maximal coupling strength is fixed at $g=0.01\omega_0$. (b) Maximal coupling strength is tuned from $g=0.01\omega_0$ to $g=0.03\omega_0$, and acceleration is fixed at $\mathcal{A}_0$.
Remaining parameters are $\omega_q=\omega_0=2\pi \times 4~\text{GHz}$, $k_0 = \pi/0.01~m^{-1}$, $T_1 = 10~\mu s$, $T_2= 20~\mu s$, and $\kappa=100~\text{KHz}$.}
\end{figure}

\begin{figure*}
\includegraphics[width=1.05\linewidth]{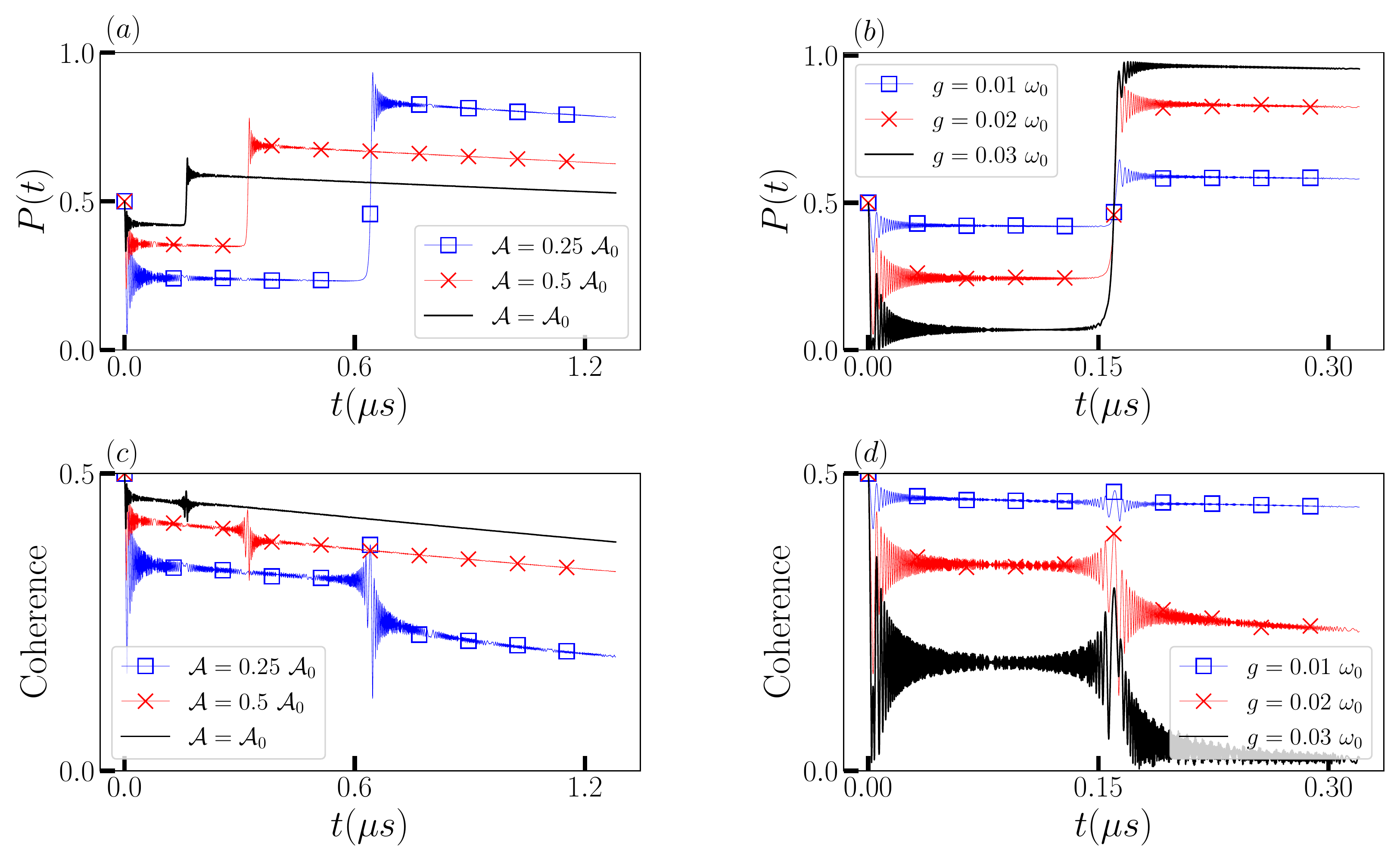}
\caption{\label{sup} Numerical simulations of qubit excitation probability and coherence with initial qubit-field state $(|G,0\rangle + |E,0\rangle)/\sqrt{2}$. (a), (c) Qubit dynamics for different values of acceleration with a fixed coupling strength $g=0.01\omega_0$. (b), (d) Qubit dynamics for different values of coupling strength with a fixed acceleration $\mathcal{A}= 10^{15}m/s^2$. Remaining parameters are $\mathcal{A}_0 = 10^{15}m/s^2$, $\omega_q=\omega_0=2\pi \times 4~\text{GHz}$, $k_0 = \pi/0.01~m^{-1}$, $T_1 = 10~\mu s$, $T_2= 20~\mu s$, and $\kappa=100~\text{KHz}$.}
\end{figure*}

Let us begin by modeling a qubit moving with a uniform acceleration $\mathcal{A}$ in a cavity, as depicted in Fig.~\ref{model}(a). The qubit trajectory is given by 
\begin{equation}
x_q(t)=\frac{c}{\mathcal{A}}\sqrt{\mathcal{A}^2t^2+c^2}-\frac{c^2}{\mathcal{A}},\label{acc}
\end{equation} where $c$ is the speed of light in vacuum.
The total qubit-field Hamiltonian can be written as (by letting $\hbar =1$ for simplicity)
\begin{align}
H &= \frac{\omega_q}{2}\sigma_z + \omega_0 a^\dagger a +H_I[x_q(t)]  \nonumber\\
\text{with}\quad H_I[x_q(t)] &= g \cos[k_0 x_q(t)]\sigma_x (a^\dagger+a), 
\label{eq:SM}
\end{align}
where $\omega_q$ is the transition frequency between the qubit ground state $|G\rangle$ and the excited state $|E\rangle$; $\sigma_z$ and $\sigma_x$ are Pauli matrices acting on the qubit; and $a$ ($a^\dagger$) is the annihilation (creation) operator of the cavity field. 
Here, we assume that the qubit effectively interacts with the fundamental cavity field mode at frequency $\omega_0=vk_0$ and wave number $k_0 = \pi/L$, where $v$ is the phase velocity of the field mode, and $L$ is the cavity length. The motion of the qubit $x_q(t)$ is encoded in the interaction Hamiltonian $H_I[x_q(t)]$ with the maximal coupling strength $g$.

According to Refs.~\cite{felicetti2015,Srinivasan2011}, the system can be simulated using a superconducting circuit, as shown in Fig.~\ref{model}(b). The circuit is composed of two capacitively shunted SQUID loops that couple to a LC resonator. Theoretical investigation~\cite{Gambetta2011} revealed that these two coupled SQUID loops can be regarded as an effective qubit, where the qubit transition frequency and the qubit-resonator coupling strength can be tuned via modulation of the external fluxes $\Phi_{x}^a$ and $\Phi_{x}^b$ flowing through the SQUID loops. Moreover, it is possible to modulate the coupling strength while the qubit transition frequency is maintained a constant. In this case, the qubit-resonator interaction Hamiltonian can be modeled as
\begin{equation}
\tilde{H}_I(\Phi_{x}^a,\Phi_{x}^b) = \tilde{g}(\Phi_{x}^a,\Phi_{x}^b)\sigma_x(a^\dagger +a),
\end{equation}
where the dependence of the coupling strength $\tilde{g}$ on these two fluxes, $\Phi_{x}^a$ and $\Phi_{x}^b$, is further explained in Ref.~\cite{Gambetta2011}. Accordingly, the qubit motion depicted in Eq.~\eqref{eq:SM} can be simulated using this circuit through the modulation of the external fluxes, such that 
\begin{equation}
\tilde{g}[\Phi_x^a(t),\Phi_x^b(t)] = g\cos[k_0 x_q(t)].
\end{equation}
For this particular design, the maximally reachable simulated acceleration is approximately $10^{15} m/s^2$ (see Ref. \cite{felicetti2015} and the references therein).

Note that there is a built-in symmetry for the Hamiltonian in Eq.~\eqref{eq:SM}~\cite{PhysRevA.99.052328}, such that the trajectory given in Eq.~\eqref{acc} is equivalent to a trajectory wherein the qubit bounces back and forth between two ends of the cavity. Therefore, in a simulation scenario, the symmetry prevents the qubit from leaking out of the cavity. 

The time evolution is modeled by the following master equation, which can be numerically solved using the QuTip open-source python package~\cite{johansson2012,johansson2013}:
\begin{align}
&\frac{\partial{\rho}}{\partial{t}} = -i[H,\rho]+\kappa \mathcal{D}[a]\rho+\Gamma \mathcal{D}[\sigma_-]\rho+\Gamma_\phi \mathcal{D}[\sigma_z]\rho \nonumber\\
&\text{with}~\mathcal{D}[O]\rho = \frac{1}{2}(2O\rho O^\dagger-\rho O^\dagger O-O^\dagger O\rho).\label{meq}
\end{align}
Here, $\kappa$ is the photon decay rate, $\Gamma=1/T_1$ the qubit decay rate, and $\Gamma_\phi=(1/T_2-1/2T_1)/2$ the qubit dephasing rate, where $T_1$ and $T_2$ are the qubit relaxation time and coherence time, respectively. 
We use the typical circuit-QED parameters $(\omega_q=\omega_0=2\pi \times 4~\text{GHz}$, $k_0 = \pi/0.01~m^{-1}$, $T_1 = 10~\mu s$, $T_2= 20~\mu s$, and $\kappa=100~\text{KHz}$), which are achievable in current experiments~\cite{zhang2017suppression}. The maximal coupling strength $g$ applied in this study is below the ultrastrong coupling regime ($ g < 0.1\omega_0$)~\cite{kockum2019ultrastrong}. Furthermore, we consider a five-dimensional Fock space for the field mode, where convergent numerical results can be obtained.

In Fig.~\ref{radiation_p}, we plot numerical simulations of the qubit excitation probability $P(t) = \langle E|\rho(t) |E \rangle$ at different accelerations and coupling strengths. Here, we consider the qubit as being initialized in its ground states $|G\rangle$. Moreover, the field is initialized in the vacuum state $|0\rangle$. An interesting feature is that, after a period of time, $P(t)$ will be significantly enhanced and reach an equilibrium value (before the qubit decay, which results from the dissipation). Either a decrease in the acceleration or an increase in the coupling strength also leads to enhancement of the Cherenkov radiation. In addition, the moment at which enhancement starts to occur depends on the acceleration magnitude. Because there is initially no excitation in the entire system, the enhancement must originate from the activation of counter-rotating transition, which simultaneously excites the qubit and emits a photon. Such an enhancement has also been identified as a cavity-enhanced Unruh effect~\cite{scully2003,hu2004comment,scully2004scully,belyanin2006quantum,PhysRevA.99.053833,PhysRevLett.103.147003}, wherein the counter-rotating transition can be activated via movement of the qubits without requiring ultrastrong coupling between the qubits and the cavity field.

To gain further insight, we calculate the counter-rotating photon emission probability by using the standard perturbation technique. Note that the threshold time $t^*$, where the enhancement starts to occur, is small enough ($\frac{\mathcal{A}t^*}{c}<1$) that a non-relativistic limit can be obtained, such that
\begin{equation}
x_q(t)\simeq \frac{1}{2}\mathcal{A}t^2.
\end{equation}
With this approximation, the leading order of the transition probability from $|G,0\rangle$ to $|E,1\rangle$ at time $t$ can be written as
\begin{eqnarray}
P(t) &=&  \biggl| \int_0^{t}~d\tau~ g e^{i(\omega_q+\omega_0)\tau}\cos(k_0\frac{1}{2}\mathcal{A}\tau^2) \biggl|^2 \nonumber \\ 
&=& \frac{g^2\pi}{8\mathcal{A}k_0}\biggl|~e^{i\frac{(\omega_0+\omega_q)^2}{\mathcal{A}k_0}}\text{erf}\Big [ \frac{1+i}{2\sqrt{\mathcal{A}k_0}}(-\mathcal{A}k_0 \tau +\omega_0 +\omega_q) \Big]\Big| ^t_0 \nonumber\\
&&~~~~~~~~~-\text{erfi}\Big [ \frac{1+i}{2\sqrt{\mathcal{A}k_0}}(\mathcal{A}k_0 \tau +\omega_0 +\omega_q) \Big]\Big| ^t_0 ~\biggl|^2.
\label{transition P}
\end{eqnarray}
We can infer that the transition will be significantly enhanced right after the following relation is satisfied:
\begin{equation}
-\mathcal{A}k_0t^*+\omega_0 + \omega_q =0,
\end{equation}
or
\begin{equation}
v_c=\mathcal{A}t^*=\frac{\omega_0 + \omega_q}{k_0}. \label{threshold}
\end{equation}

Because $\mathcal{A}t$ can be identified as the non-relativistic velocity of the qubit, Eq.~\eqref{threshold} suggests that the requirement for transition enhancement is that the qubit velocity must be higher than the threshold velocity $v_c$. This result is also manifested in Ref.~\cite{sabin2017}, where $v_c$ has been identified as the Cherenkov threshold. Accordingly, the enhancement of the transition can be interpreted as the Cherenkov radiation triggered by the accelerating qubit. 

As shown in Eq.~\eqref{transition P}, the magnitude of the transition depends on $g^2/\mathcal{A}$; that is, either an increase in $g$ or a decrease in $\mathcal{A}$ results in enhancement of the Cherenkov radiation. It is somewhat intuitive that for cavity QED systems, increasing the qubit-field coupling strength also enhances the counter-rotating transitions. On the other hand, however, enhancement of the Cherenkov effect due to the acceleration $\mathcal{A}$ is somewhat counter-intuitive, because from the viewpoint of the Unruh effect, the qubit experiences a hotter temperature, or more excitation, when its acceleration is increased. The prediction in Eq.~\eqref{transition P} shows an opposite result, because decreasing the acceleration results in enhancement of the excitation, i.e., the Cherenkov radiation. In fact, several researches have pointed out that the accelerated particle coupled to the vacuum can be cooled down as the acceleration is increased, and this phenomenon is termed as anti-Unruh effect~\cite{brenna2016anti,PhysRevD.94.104048,liu2016decoherence,PhysRevD.97.045005}.

In contrast with the previous set-up, we now set the qubit initial state to be the superposition state $(|G\rangle + |E\rangle)/\sqrt{2}$, whereas the field is still initialized in the vacuum state $|0\rangle$. In Fig.~\ref{sup}, we plot the time evolutions of the qubit excitation probability and qubit coherence $|\langle E|\rho(t)|G\rangle|$ for different coupling strengths $g$ and accelerations $\mathcal{A}$. We can observe fast transients around $t=0$. It could be resulted from the non-adiabatic effect, since at the moment $t=0$ the qubit-field interaction is suddenly turned on. The physical intuition behind the non-adiabatic effect is that the sudden change of the interaction kicks the system and results in drastic transients for system dynamics. When the coupling strength or the acceleration is increased, the kick (sudden change of the interaction) becomes stronger such that one would expect more drastic transients. However, the result of the increased acceleration does not follow the intuition, because the transients’ magnitude is actually smaller when the acceleration is increased. We therefore point out that it is another counter-intuitive aspect of qubit acceleration. Moreover, the enhancement of the qubit excitation probability accompanies the suppression of the coherence after the Cherenkov threshold is crossed, where the degree of the qubit excitation probability enhancement and the coherence suppression are enhanced by either a decrease in $\mathcal{A}$ or increase in $g$. Note that the coherence remains at zero for the entire time evolution when the qubit is initialized in $|G\rangle$.

With the qubit dynamics for the two different initial states, we can infer how well, depending on different $g$ and $\mathcal{A}$, the qubit is able to transmit quantum information after crossing the threshold. According to the concept of the memory effect~\cite{PhysRevLett.103.210401,RevModPhys.89.015001}, the loss of quantum information due to a given dynamical map $\mathcal{N}_t$ can be quantified by the decrease in the trace distance for a given pair of initial states after being sent into the map. To be more specific, consider two initial states, $\rho_1(0)$ and $\rho_2(0)$, with non-zero trace distance. The reduction of the trace distance between the states after insertion into the dynamical map, i.e., $\mathcal{N}_t[\rho_1(0)]$ and $\mathcal{N}_t[\rho_2(0)]$, can be regarded as loss of the quantum information. Accordingly, because the qubit approaches $|E\rangle$ for the two different initial conditions when $g$ is increased or $\mathcal{A}$ is decreased, it is implied that modulation of $g$ and $\mathcal{A}$ reduces the trace distance of the dynamics for the different initial states, and therefore enhances quantum information leakage to the environment. 

Motivated by this observation, in the upcoming parts of the research, we employ measurements of the quantum direct cause, which can be used to quantify the upper bound of the channel capacity, that is, the ability of the qubit to propagate quantum information for a given time.

\section{Measurement of causal influence using temporal correlations \label{causal_measure}}

\begin{figure*}
\includegraphics[width=0.85\paperwidth]{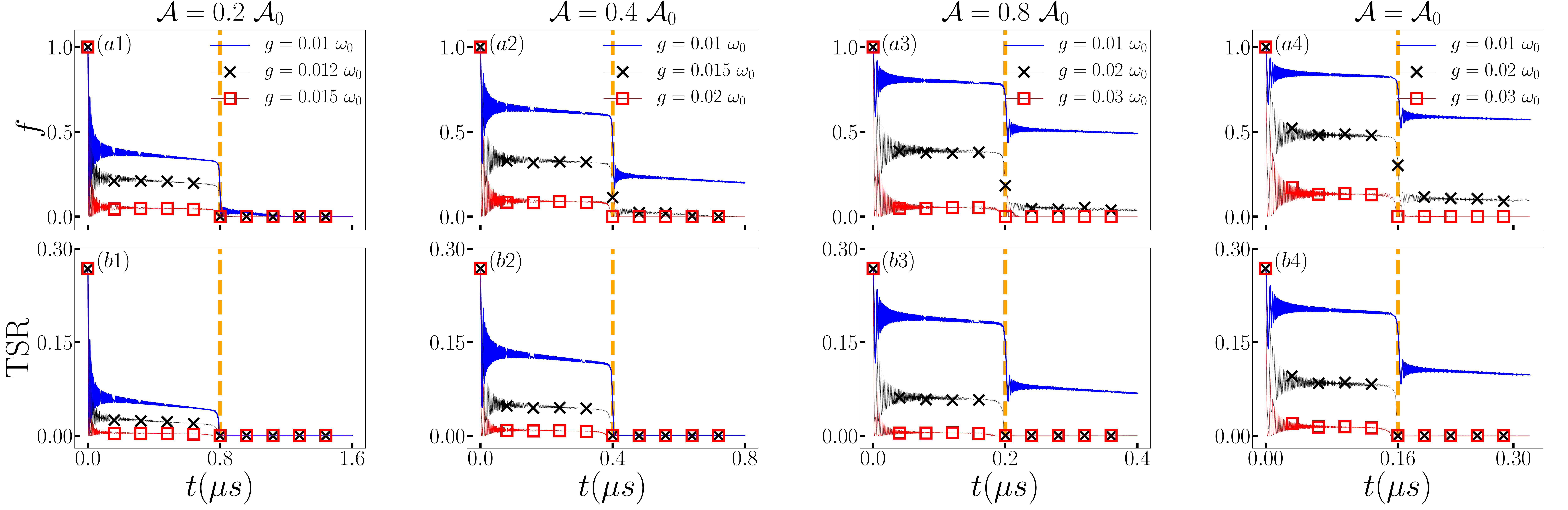}
\caption{\label{TSR_F_single_mode} Evolutions of $f$-function (a1--a4) and TSR (b1--b4) for different qubit accelerations $\mathcal{A}$ and maximum coupling strengths $g$. Remaining parameters are $\omega_q=\omega_0=2\pi \times 4~\text{GHz}$, $k_0 = \pi/0.01~m^{-1}$, $T_1 = 10~\mu s$, $T_2= 20~\mu s$, and $\kappa=100~\text{KHz}$. Orange dashed lines denote Cherenkov thresholds for different acceleration values predicted using Eq.~\eqref{threshold}}
\label{cau_T}
\end{figure*}

\begin{figure}
\includegraphics[width=1\linewidth]{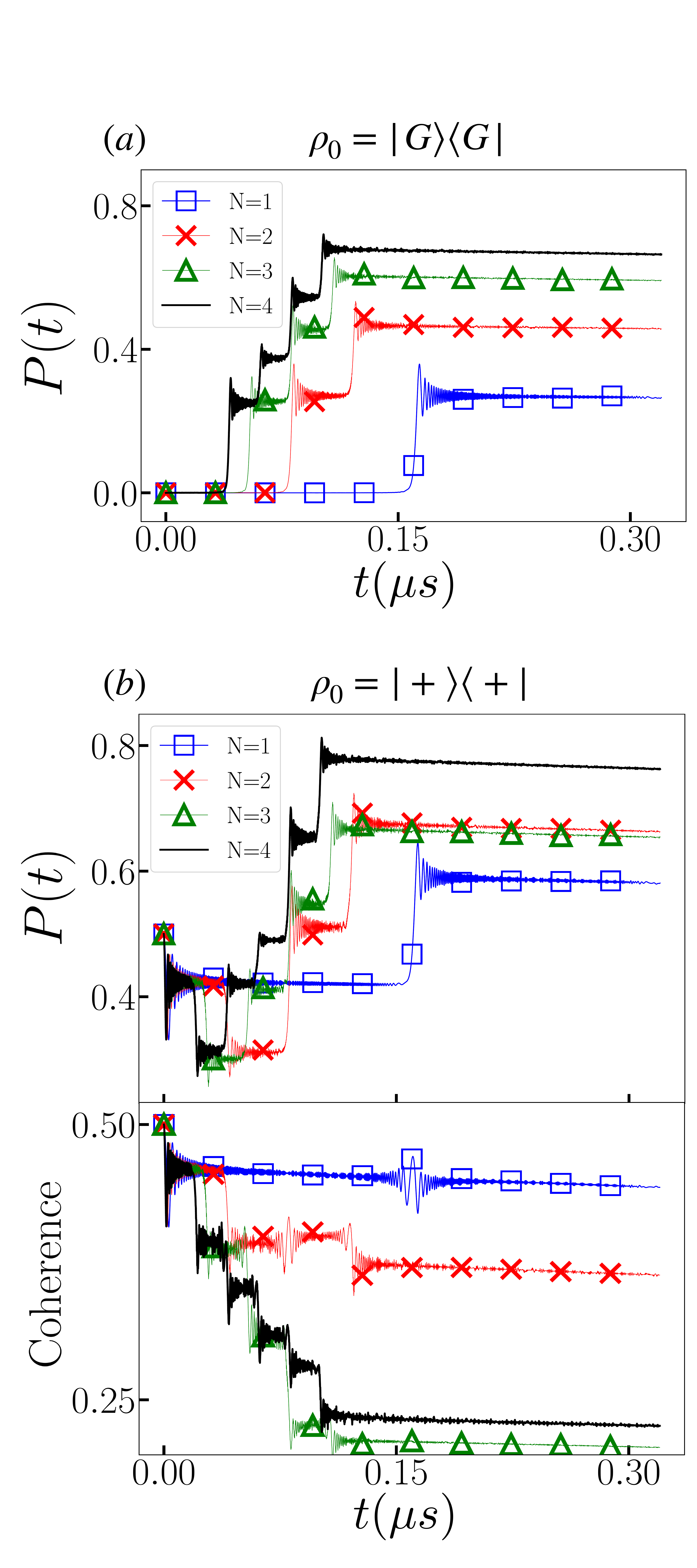}
\caption{\label{mt_te} Time evolutions of qubit excitation probability and coherence for different numbers of field modes $N$ and for different initial states, $|G\rangle$ and $|+\rangle = (|G\rangle+|E\rangle)/\sqrt{2}$.
Here, qubit acceleration is fixed at $\mathcal{A}=10^{15}\text{m}/\text{s}^2$ and  $g_0 = 0.01\omega_0$. Remaining parameters are $\omega_q=\omega_0=2\pi \times 4~\text{GHz}$, $k_0 = \pi/0.01~m^{-1}$, $T_1 = 10~\mu s$, $T_2= 20~\mu s$, and $\kappa=100~\text{KHz}$.}
\end{figure}

\begin{figure}
\includegraphics[width=1\linewidth]{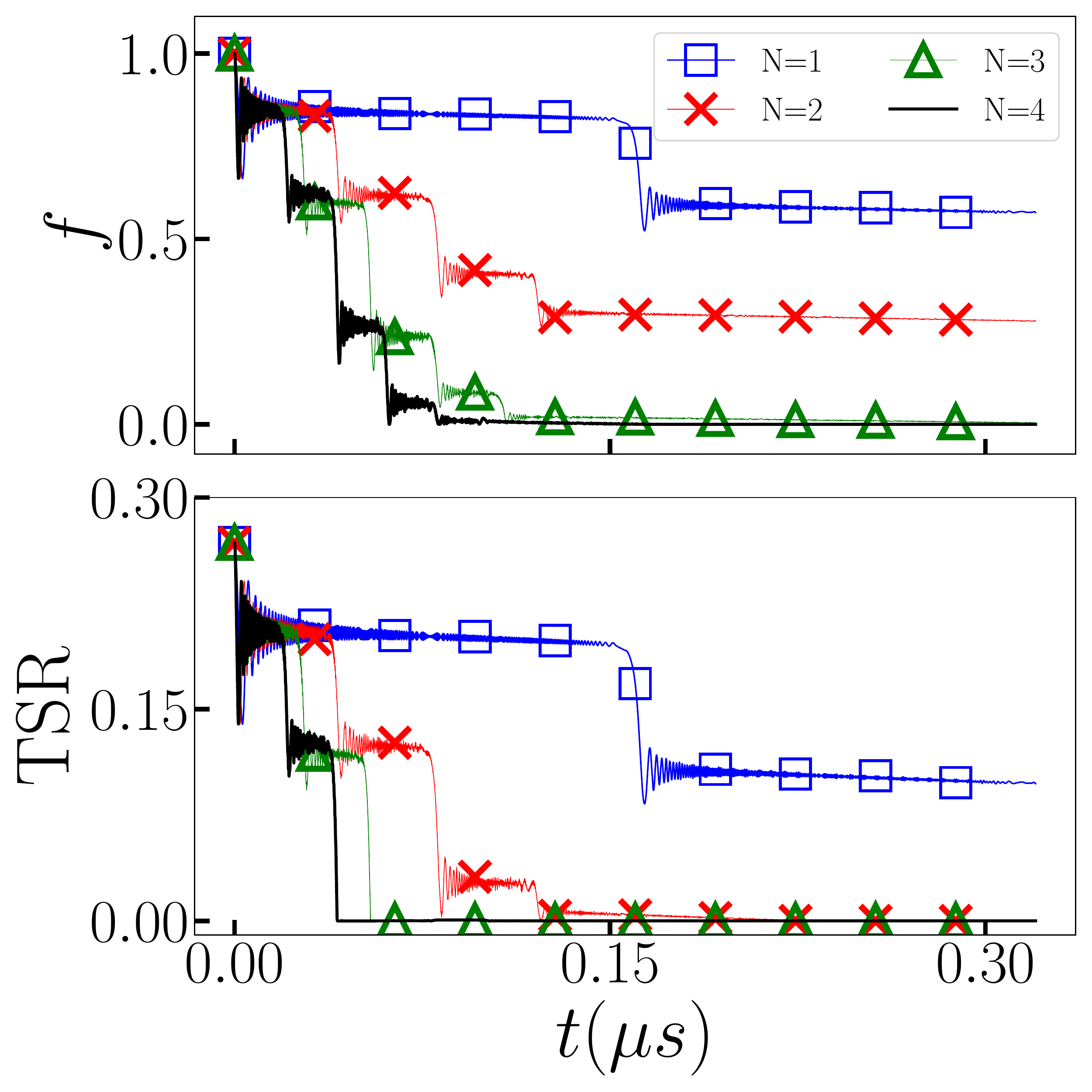}
\caption{\label{mt} TSR and $f$-function for different numbers of field modes $N$ with fixed qubit acceleration $\mathcal{A}=10^{15}\text{m}/\text{s}^2$ and  $g_0 = 0.01\omega_0$. Remaining parameters are $\omega_q=\omega_0=2\pi \times 4~\text{GHz}$, $k_0 = \pi/0.01~m^{-1}$, $T_1 = 10~\mu s$, $T_2= 20~\mu s$, and $\kappa=100~\text{KHz}$.}
\end{figure}

Here, we give a short review on measurements of quantum direct cause based on PDM and TS, which are applicable to the current model.

Let us start from the PDM formulation. A PDM is constructed from generalized quantum state tomography via the collection of measurement outcomes from a single system at two successive times. By definition, for a single qubit system, the corresponding PDM~\cite{fitzsimons2015} is written as
\begin{equation}
R = \sum_{i,j=0}^3\langle\{\sigma_i^{t_0},\sigma_j^{t_f}\}\rangle \sigma_i \otimes \sigma_j,
\end{equation}
where $\langle\{\sigma_i,\sigma_j\}\rangle$ is the expectation value of the product of the outcome of measuring $\sigma_i$ at initial time $t_0$ and the outcome of measuring $\sigma_j$ at later time $t_f$. The Hilbert space for $R$ is $\mathcal{H}_A^{t_0}\otimes\mathcal{H}_A^{t_f}$, which is the tensor product of the state space of the qubit at initial time and later time. A PDM shares many similarities with a bipartite density matrix except that it is not necessary to be positive-semidefinite. Because any negative $R$ cannot be reinterpreted as a regular bipartite quantum state, it can eliminate common-causal explanations for the correlations between two qubits. Based on such insight, a measure of the quantum direct cause (or the direct-causal influence), referred to as an $f$-function, has been proposed ~\cite{fitzsimons2015}: 
\begin{equation}
f = \sum_i |\mu_i |,
\end{equation}
which is the summation of all negative eigenvalues $\mu_i$ for a given $R$. It appears that the negativity of a PDM has an operational meaning for quantum communication. According to Pisarczyk \textit{et al.}~\cite{PhysRevLett.123.150502}, the negativity (in this research, the logarithmic negativity) serves as a computable upper bound of channel capacity for a quantum channel, implying that the degrees of quantum direct cause characterize the ability of the channel to transmit quantum information. Accordingly, in this study, measuring the direct-causal influence is equivalent to quantifying the capability of the accelerating qubit for quantum information propagation.

Let us now turn to the temporal steering scenario, where a system is measured at initial time $t_0$, and the resulting dynamics are collapsed (steered) into different states at later time $t_f$. Operationally, the key quantity of interest in TS is characterized by a set of unnormalized states referred to as a temporal steering assemblage: $\{\sigma_{a|x}(t) = p(a|x)\rho_{a|x}(t)\}_{a,x}$, where $p(a|x)$ is the probability of obtaining outcome $a$ conditioned on the measurement choice $x$ performed at $t_0$, and $\rho_{a|x}(t)$ is the conditional quantum state at time $t$, where the steered evolution is characterized. For simplicity, we consider that the measurements are characterized by the eigenoperators of three Pauli matrices, i.e., $\{\sigma_x,\sigma_y,\sigma_z\}$. We highlight here that the element in the assemblage can also be derived from PDM using the following equation~\cite{ku2018}: 
\begin{equation}
\sigma_{a|x}(t) = \tr_{t_0}[R(E_{a|x}\otimes \mathbb{I})], \label{Born}
\end{equation}
where the set $E_{a|x}$ denotes the measurements conducted at $t_0$. Furthermore, the hierarchical relation between PDM and TS can be obtained from Eq.~\eqref{Born}, indicating that the TS can be regarded as a weaker measure of direct-causal influence. The magnitude of temporal steerability can be quantified by several distant measures between the given assemblage and what can be classically interpreted as what is known as the local hidden state (LHS) model, which is written as  
\begin{equation}
\sigma_{a|x}^{\text{LHS}}(t) = \sum_\lambda p(\lambda)p(a|x,\lambda) \rho_\lambda,
\end{equation} where $\{p(\lambda),\rho_\lambda\}$ is an ensemble of ontic states  distributed with the random variable $\lambda$, and $\{p(a|x,\lambda)\}$ denotes classical post-processing. In this research, we utilize the measure of steerability known as the temporal steering robustness (TSR), which is defined as~\cite{ku2016,ku2018}
\begin{align}
&\text{TSR}[\sigma_{a|x}(t)] = \min~\alpha \nonumber\\
&\text{s.t.}~~\frac{1}{1+\alpha}\sigma_{a|x}(t) + \frac{\alpha}{1+\alpha}\tau_{a|x}=\sigma^{\text{LHS}}_{a|x}~~\forall a,x,
\end{align}
where $\tau_{a|x}$ is an arbitrary noisy assemblage element. This value quantifies the minimal noise required to destroy the steerability for a given assemblage.

Note that, as explained in the following section, we impose another restriction, referred to as no-signaling in time condition~\cite{PhysRevA.93.022123,ku2018}, which can be realized via initialization of the qubit in the maximally mixed state $\rho_0 = \mathbb{I}/2$. In this case, we can negate the possibility of a \textit{classical} direct-causal effect, e.g., influence from some hidden communication channels.

\section{Results and discussions\label{numerical_res}}

In this section, to reveal the effect of the Cherenkov threshold, we present the numerical results of utilizing the measures of quantum direct cause introduced from the previous sections. In Fig.~\ref{cau_T}, we plot the TSR and the $f$-function as functions of time with respect to different accelerations and coupling strengths. In general, the quantum direct cause will decrease sharply when the threshold is crossed. In addition, the results suggest that either a decrease in the acceleration $\mathcal{A}$ or an increase in the maximal coupling strength $g$ can suppress and eventually eliminate the remaining direct-causal influence. This result is consistent with previous observations shown in Fig.~\ref{radiation_p} and Fig.~\ref{sup}, where the equilibrium states of the qubit can approach the excited state $|E\rangle$ when the values of $g$ and $\mathcal{A}$ are changed.

Here, one can infer that either a decrease in the acceleration to $\mathcal{A}=2\times10^{14}~m/s^2$ with a fixed maximal coupling strength $g=0.01\omega_0$, or an increase in the maximal coupling strength to $g=0.03\omega_0$ with a fixed acceleration $\mathcal{A}=10^{15}~m/s^2$, can result in the suppression of the direct-causal influence to zero. Therefore, for these cases, where the remaining direct-causal influence can be fully eliminated, the Cherenkov threshold serves as the speed limit for the qubit to transmit quantum information. Note that for some parameters of $g$ and $\mathcal{A}$, TSR drops to zero, whereas the $f$-function does not. The difference results from the hierarchical relation between these two measures~\cite{ku2018}.

In previous sections, we focus on the single-mode model. However, to simulate a genuine cavity, which can support many field modes, we should consider a multi-mode model, which is generally described by the following Hamiltonian:
\begin{equation}
H = \frac{\omega_q}{2}\sigma_z + \sum_{n=0}^{N-1} \omega_n a_n^\dagger a_n + g_n \cos(\frac{1}{2}k_n\mathcal{A}t^2) \sigma_x(a_n^\dagger + a_n),
\end{equation}
where $N$ is the number of the field modes involved in the cavity. $\omega_n = (n+1)\omega_0$, $g_n = \sqrt{n+1} g_0$, and $k_n = (n+1)k_0$ are the frequency, coupling strength, and wave vector, respectively, for each field mode $n$. The dynamic is described by the master equation when $\kappa\mathcal{D}[a]\rho$ is replaced with $\kappa\sum_{n=0}^{N-1}\mathcal{D}[a_n]\rho$ in Eq.~\eqref{meq}. Moreover, we consider a five-dimensional Fock space for each field mode, which also produces convergent results.

In Fig.~\ref{mt_te}, we plot the qubit dynamics for different numbers of field modes and for the two aforementioned qubit initial states, $|G\rangle$ and $(|G\rangle+|E\rangle)/\sqrt{2}$. The acceleration $\mathcal{A}$ and the coupling strength $g_0$ are fixed at $10^{15}m/s^2$ and $0.01\omega_0$, respectively. We can infer that several transitions occur for the multi-mode model, where the number of transitions depends on the number of modes. When the qubit is initialized in $|G\rangle$, the excited state population is enhanced after each transition. In addition, the equilibrium value of the excited state population increases when the number of modes is increased. Note that the coherence remains at zero throughout the evolution. However, when the qubit is initialized in $(|G\rangle+|E\rangle)/\sqrt{2}$, the transitions do not always lead to enhancement of the qubit excitation probability. We can observe that the qubit excitation probability decreases after the first transition, and then again increases for the following transitions. Additionally, increasing the number of modes does not always increase (decrease) the equilibrium qubit excitation probability (coherence). The equilibrium value of the qubit excitation probability for a two-mode case is larger than that for a three-mode case. Similarly, the equilibrium value of the coherence for a three-mode case is smaller than that for a four-mode case. On the other hand, we can still observe that the coherence decreases for each transition. From the viewpoint of an open quantum system, decreases in the coherence imply the existence of strong system--environment coupling, which leads to the leakage of quantum information. However, because the magnitude of the coherence is not a proper quantifier of quantum direct cause, we should again employ the aforementioned measures to produce a quantitative result.

The numerical results of the quantum direct cause for the multi-mode model are presented in Fig.~\ref{mt}. We can clearly determine that the direct-causal influence decreases after each transition. When the number of field modes increases, the equilibrium value of the direct-causal influence also decreases and, eventually, vanishes from the addition of a sufficient number of modes. To eliminate the direct-causal influence, two modes are necessary for TSR, whereas three modes are required for the $f$-function. Such a difference can again be seen as a manifestation of the hierarchical relation between TSR and $f$-function~\cite{ku2018}.

There are several studies \cite{munoz2018,jonsson2014,sabin2011,zohar2011} that have already revealed that introducing additional field modes can suppress the signaling velocity in a Fermi's test. However, the physical origin of the suppression in their cases is different from that in our scenario. In their cases, the information is propagated by flying photons. The suppression of the signaling velocity is due to the localization of the photon wave packet, because it is more likely for the photon to form a localized wave packet when a sufficient number of field modes is available in the system. In our model, on the other hand, because the quantum information is carried by an accelerating qubit, the cavity field should be treated as the environment to the qubit. From the perspective of an open quantum system, the information stored in the system tends to leak out to the environment when the size (degrees of freedom) of the environment is much larger than the system. Moreover, increasing the number of field modes increases the size of the environment. The numerical results suggest that this also magnifies the tendency of the qubit to lose quantum information and results in further suppression of the quantum direct cause.

A few remarks on the multi-mode model should be given here. In general, the couplings of an artificial atom to a field are not restricted in the lowest two levels. Transitions beyond the lowest two levels are possible. In this case, we shall still consider the lowest two levels as our system. From the viewpoint of an open quantum system, the other levels are regarded as another reservoir. Therefore, transitions to other levels may lead to information leakage, and we can expect the quantum direct-causal influence to possibly suffer additional suppression. 

\section{Summary}
In summary, we investigated how to quantitatively measure the quantum direct cause for a qubit accelerating across the Cherenkov threshold. Conceptually, the system does not involve two different parties, acting as signal sender and receiver, to allow a Fermi's test to be performed \cite{munoz2018,jonsson2014,sabin2011,zohar2011}. In addition, the system is characterized by a time-dependent Hamiltonian, which forbids us from analyzing the system with its energy spectrum \cite{munoz2018}. These two aspects motivated us to investigate causation on a single moving system using recently developed direct-causal measures based on temporal quantum correlations. 

We numerically demonstrated that either an increase in the qubit-field coupling strength or decrease in the qubit acceleration can result in the suppression of quantum direct cause after the threshold crossing. By observing the qubit dynamics for different initial states, we can infer that modulation of the coupling strength and acceleration causes the qubit to approach the excited state after crossing the Cherenkov threshold. This makes it more difficult to retrieve the quantum information, which is initially encoded in the qubit.

To simulate a genuine cavity, we extended our investigation to a multi-mode model. We determined that the qubit dynamics involves several transitions. Moreover, each transition always occurred together with a decrease in the qubit coherence, implying the existence of strong system--environment coupling, which results in the leakage of quantum information. We again employed measurements of quantum direct cause for the multi-mode model. The results suggest that increasing the number of modes can further suppress direct-causal influence. The suppression is an effect of the increase in system--reservoir coupling due to the enlarged size of the environment.


\begin{acknowledgments}
We thank Neill Lambert and Roberto Stassi for insightful discussion. This work is supported partially by the National Center for Theoretical Sciences and Ministry of Science and Technology, Taiwan, Grants No. MOST 107-2628-M-006-002-MY3, and MOST 108-2627-E-006-001, and the Army Research Office (under Grant No. W911NF-19-1-0081).
\end{acknowledgments}


%

\end{document}